\documentclass[12pt,rmp,twocolumn,natbib,showkeys,pre,floatfix]{revtex4}

\usepackage[dvips]{graphicx}

\begin{document}

\title{Dynamics of a FitzHugh-Nagumo system subjected to autocorrelated noise}

\author{\firstname{D.} \surname{Valenti\footnote{e-mail: valentid@gip.dft.unipa.it}}}
\author{\firstname{G.} \surname{Augello\footnote{e-mail: augello@gip.dft.unipa.it}}}
\author{\firstname{B.} \surname{Spagnolo\footnote{e-mail: spagnolo@unipa.it}}}
\affiliation{Dipartimento di Fisica e Tecnologie Relative and
CNISM-INFM,
\\ Group of Interdisciplinary
Physics\footnote{http://gip.dft.unipa.it}, Universit$\grave{a}$ di
Palermo, \\ Viale delle Scienze, I-90128, Palermo, Italy}

\begin{abstract}

We analyze the dynamics of the FitzHugh-Nagumo (FHN) model in the
presence of colored noise and a periodic signal. Two cases are
considered: (i) the dynamics of the membrane potential is affected
by the noise, (ii) the slow dynamics of the recovery variable is
subject to noise. We investigate the role of the colored noise on
the neuron dynamics by the mean response time (MRT) of the neuron.
We find meaningful modifications of the resonant activation (RA) and
noise enhanced stability (NES) phenomena due to the correlation time
of the noise. For strongly correlated noise we observe suppression
of NES effect and persistence of RA phenomenon, with an efficiency
enhancement of the neuronal response. Finally we show that the
self-correlation of the colored noise causes a reduction of the
effective noise intensity, which appears as a rescaling of the
fluctuations affecting the FHN system.

\end{abstract}

\date{\today}


\maketitle

\section{Introduction}

Physical and biological systems are continuously perturbed by random
fluctuations produced by noise sources always present in open
systems~\cite{Complex}. In physical and biological systems, noise
can be responsible for several interesting and counterintuitive
effects. ~Resonant activation (RA) and noise enhanced stability
(NES) are two noise activated phenomena that have been widely
investigated in a dissipative optical lattice, spin systems,
Josephson junctions, chemical and biological complex systems, and
financial markets~\cite{RA,NES,Nes-theory,SR}. Because of the
stochastic resonance phenomenon, a weak electric stimulus can be
enhanced up to
the detection threshold of the neural system~\cite{Wiesenfeld}.\\
\indent Recently the increasing interest in neuronal dynamics gave
rise to many studies in this field. Two approaches, widely used to
investigate the neuron response to an external stimulus, are the
Hodgkin-Huxley model~\cite{Hodgkin} and its simplified version, the
FitzHugh-Nagumo (FHN)
model~\cite{Bonhoeffer,van_der_Pol,FitzHugh,Hale,Rocsoreanu,Nagumo,Keener}.
A FHN system, in the presence of fluctuations, because of its
intrinsic nonlinearity, can give rise to interesting noise induced
phenomena: modification of detection threshold by manipulation of
noisy parameters~\cite{Pei}, noise-induced activation and coherence
resonance~\cite{Pikovsky}, stochastic resonance in the presence of
colored noise with a $1/f^\beta$ spectrum ($0 \leq \beta \leq
2$)~\cite{Nozaki}, optimization of the interspike time and
dependence of neuron firing on subthreshold periodic signal and
additive noise~\cite{Longtin}, resonant activation and
noise enhanced stability~\cite{Pankratova}.\\
\indent Despite the broad interest in FHN system, however,
understanding the role of the fluctuations and their correlation in
nerve cells still remain elusive. This is why it is of particular
importance the investigation of the mean response time of a single
neuron-like element as a function of the noise parameters in the
presence of various external perturbations.\\
In this paper, by using a colored noise given by the archetypal
Ornstein-Uhlenbeck process, with correlation time $\tau$, we study
the role of the fluctuations on the FHN system. In particular we
analyze the mean response time (MRT) as a function both of the noise
intensity and the correlation time and we find that the RA and NES
phenomena observed in the presence of white noise~\cite{Pankratova}
show meaningful modifications when a colored noise source is used.
 The paper is organized as follows. In the first paragraph we
 shortly introduce the Hodgkin-Huxley (HH) system, the archetypal model to describe
neuronal dynamics, and the FitzHugh-Nagumo (FHN) system, which is a
simplified version of the HH model, discussing briefly the
motivations for using, instead of the more realistic HH approach,
the FHN model. In order to analyze the dynamics of a single neuron,
we take into account a FHN system driven by a periodic signal and we
analyze the deterministic dynamics. Afterwards we investigate the
stochastic FHN system. In particular we consider the effects of the
fluctuations by inserting in the deterministic FHN model a colored
noise source. We analyze two different cases: the noise affects the
dynamics of the membrane potential (case I), the recovery variable
is subject to noise (case II). We introduce the mean response time
(MRT) in order to calculate the response time of the neuron, that is
the time that the membrane potential takes to reach a given
threshold. Finally we show that the self-correlation of the colored
noise causes a reduction of the effective noise intensity, that is a
rescaling effect of the fluctuations affecting the FHN system. In
the last section we draw the conclusions.

\label{sec:intro}

\section{Hodgkin-Huxley model} \label{subsec:hodgkin} In the year
1952 Alan Lloyd Hodgkin and Andrew Huxley proposed a neuronal model
to describe the dynamics of an action potential. This consists in a
"spike" of electrical discharge generated as response to an external
stimulus and travelling along the membrane of a cell. The
Hodgkin-Huxley (HH) model, which consists of four nonlinear ordinary
differential equations, describes the electrical characteristics of
excitable cells such as neurons and cardiac myocytes, explaining the
ionic mechanisms responsible for the appearance and propagation of
action potentials in the squid giant axon~\cite{Hodgkin}. The
semipermeable cell membrane separates the interior of the cell from
the extracellular liquid and acts as a capacitor. If an input
current $I(t)$ is injected into the cell, it may add further charge
on the capacitor, or leaks through the channels in the cell
membrane. Because of active ion transport through the cell membrane,
the ion concentration inside the cell is different from that in the
extracellular liquid. The Nernst potential generated by the
difference in ion concentration is represented by a battery.
Therefore the Hodgkin-Huxley type model represents the biophysical
characteristic of cell membranes.

The HH model consists of four equations. The first one, given by
\begin{equation}
C \frac{du}{dt} = - \sum_k I_k + I.\label{hodgkin_u}
\end{equation}
with
\begin{equation}
\sum_k I_k = g_{Na} m^3 h (u-E_{Na}) + g_K n^4 (u-E_K) + g_L
(u-E_L). \label{current_sum}
\end{equation}
provides the dynamics of the voltage \emph{u} across the membrane
and $I_k$ is the sum of the ionic currents which pass through the
cell membrane. The conservation of electric charge on a piece of
membrane implies that the applied current $I(t)$ may be split in a
capacitive current $I_c = C du/dt$ which charges the capacitor $C$
and further components $I_k$ which pass through the ion channels: a
sodium channel with index $Na$, a potassium channel with index $K$
and an unspecific leakage channel with resistance $R$. All channels
may be characterized by their resistance or, equivalently, by their
conductance. The leakage channel is described by a
voltage-independent conductance $g_L = 1/R$; the conductance of the
other ion channels is voltage and time dependent. If all channels
are open, they transmit currents with a maximum conductance $g_{Na}$
or $g_K$, respectively. The parameters $E_{Na}$, $E_K$, and $E_L$
are the reversal potentials. Reversal potentials and conductances
are empirical parameters whose values, referred to a voltage scale
where the resting potential is zero, were obtained by Hodgkin and
Huxley~\cite{Hodgkin}. The three additional variables $m$, $n$, and
$h$ describe the probabilities that the two channels are open. The
combined action of $m$ and $h$ controls the $Na^+$ channels. The
$K^+$ gates are controlled by $n$. The three variables $m$, $n$, and
$h$ are called gating variables. They evolve according to the
differential equations
\begin{eqnarray}
\frac{dm}{dt}&=&\alpha_m(u)\thinspace (1\thinspace-\thinspace
m)\thinspace - \thinspace\beta_m(u)\thinspace m\label{hodgkin_m}\\
\frac{dn}{dt}&=&\alpha_n(u)\thinspace (1\thinspace-\thinspace
n)\thinspace - \thinspace\beta_n(u)\thinspace n\label{hodgkin_n}\\
\frac{dh}{dt}&=&\alpha_h(u)\thinspace (1\thinspace-\thinspace
h)\thinspace - \thinspace\beta_h(u)\thinspace h\label{hodgkin_h}.
\end{eqnarray}
The functions $\alpha_m$, $\alpha_n$, $\alpha_h$, $\beta_m$,
$\beta_n$, $\beta_h$ are empirical functions of $u$ that have been
adjusted by Hodgkin and Huxley to fit the data of the giant axon of
the squid. The four
equations~(\ref{hodgkin_u}),(\ref{hodgkin_m})-(\ref{hodgkin_h})
define the Hodgkin-Huxley model.

\section{FitzHugh-Nagumo model} \label{sec:fitzhugh}

The FitzHugh-Nagumo (FHN) model is a simpler version of the
Hodgkin-Huxley model. In particular the FHN model takes into account
the excitable variable, that is the membrane potential, which
exhibits a fast dynamics, and the recovery variable, characterized
by a slow dynamics and responsible for the refractory behaviour of
the neuron. The reduction from four variables (HH model) to two
variables (FHN model) allows to focalize the analysis on the
properties of excitation and propagation, neglecting the specific
electrochemical properties of sodium and potassium ion flow. Thus
the neuron dynamics can be described in terms of a voltage-like
variable and a recovery variable. The former, governed by a cubic
nonlinearity, provides a regenerative self-excitation by a positive
feedback, the latter, subject to a linear dynamics, is responsible
for a slower negative feedback.

\begin{figure}
\begin{center}
\resizebox{0.90\columnwidth}{!}{\includegraphics{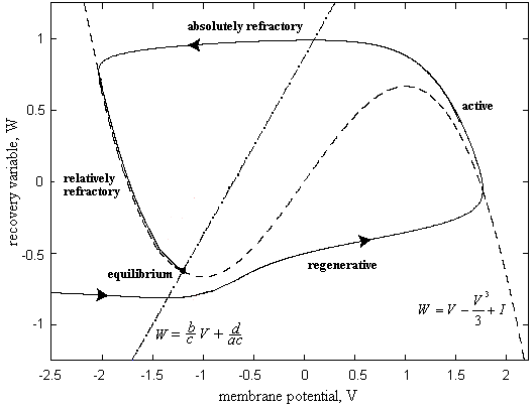}}
\caption{Phase diagram for FitzHugh-Nagumo model (modified from
FitzHugh 1961~\cite{FitzHugh}). The (V,W) trajectory (solid line)
after a spike event is in a refractory state, before reaching the
equilibrium state which corresponds to the point where the two
nullclines (dash line for $\dot{V}=0$ and dash-dot line for
$\dot{W}=0$) intersect. Here we set $b/c=1.25$, $d/(ac)=0.875$,
$I=-0.001$.\label{FHN_archetypal_phase}}
\end{center}
\end{figure}

The archetypal form for the FHN model is given by the following two
equations
\begin{eqnarray}
\dot{V} \thinspace=\thinspace f(V)\thinspace-\thinspace W\thinspace+\thinspace
I\label{archetypal_FHN_V}\\
\dot{W} \thinspace=\thinspace a\thinspace(b\thinspace
V\thinspace-\thinspace c\thinspace W)\thinspace+\thinspace
d\label{archetypal_FHN_W}
\end{eqnarray}
where $f(V)$ is a polynomial of third degree, and $a$, $b$,
$c$, $d$ are parameters with constant values.\\
Actually the Hodgkin-Huxley model is closer to the real behaviour of
neuronal dynamics. However, only projections of its four-dimensional
phase trajectories can be observed. On the other side, by using the
FitzHugh-Nagumo model, the solution for the time behaviour of the
neuronal response can be represented in a two-dimensional space.
This permits to give a geometrical explanation of important
biological phenomena connected with neuronal excitability and
spike-generating mechanism.
Eqs.(\ref{archetypal_FHN_V}),(\ref{archetypal_FHN_W}) provide the
deterministic dynamics for a FHN neuron. In fact, in the presence of
an external stimulus, with enough strong amplitude, the membrane
voltage (fast variable) increases rapidly, crossing a threshold.
This event (voltage spiking) represents the neuron firing, that is
the nerve response due to the external stimulus. After spike
generation the voltage goes rapidly under threshold and a refractory
time occurs so that during this time no further firing is possible.
This behaviour is represented in Fig.~\ref{FHN_archetypal_phase}
where a $(V,W)$ trajectory and the two nullclines
\begin{eqnarray}
W=\thinspace f(V)+\thinspace I\label{archetypal_FHN_nullcline_V}\\
W=\thinspace\frac{b}{c}\thinspace
V+\thinspace\frac{d}{ac}\label{archetypal_FHN_nullcline_W}
\end{eqnarray}
are shown for $f(V)=V-\frac{V^3}{3}$, $\frac{b}{c}=1.25$,
$\frac{d}{ac}=0.875$, $I=-0.001$.
%
%
%
\begin{figure}
\begin{center}
\resizebox{0.90\columnwidth}{!}{\includegraphics{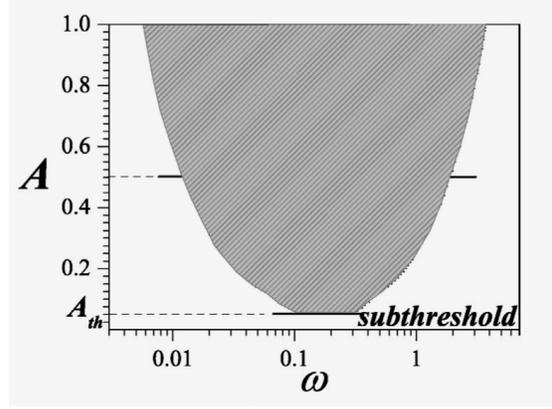}}
\caption{Parameter plane ($\omega$,$A$) for $\phi_0 = 0$.
  The grey zone corresponds to values of $\omega$ and $A$ for which deterministic firing
  occurs. The values of the parameters are $I = 1.1$, $\epsilon = 0.05$, $\varphi_0 = 0$. The
initial state is $(x_0,y_0)$ (courtesy of Pankratova et
al.~\cite{Pankratova}).\label{param_plane}}
\end{center}
\end{figure}
\section{FHN model with periodical driving signal} \label{sec:determ_model}

In the presence of a driving signal the FHN model becomes
\begin{eqnarray}
\dot{x}&=&x-\frac{x^3}{3}- y + A sin (\omega \thinspace t +
\varphi_0)
\label{FitzHugh1}\\
\dot{y}&=&\epsilon (x+I), \label{FitzHugh2}
\end{eqnarray}
where $\epsilon$ is a fixed small parameter which characterizes the
recovery process, and $A$, $\omega$ are respectively amplitude and
frequency of the external forcing.
%
\begin{figure}
\begin{center}
\resizebox{0.90\columnwidth}{!}{\includegraphics{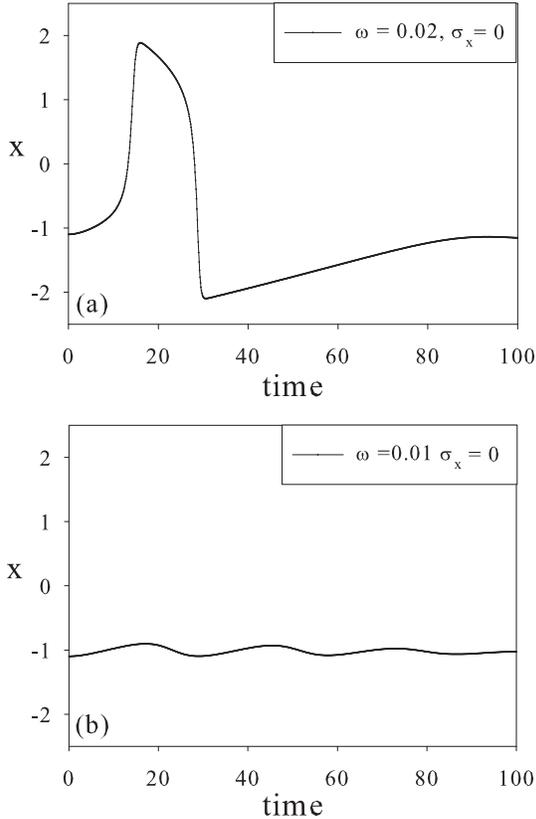}}
\caption{(a) The membrane potential $x$ exhibits a spike for $A =
0.5$ and $\omega = 0.02$ (deterministic firing region). (b) For $A =
0.5$, $\omega = 0.01$ no spike appears (deterministic no-firing
region): the values of the parameters don't allow the system to
fire. Parameter values and initial conditions are the same of
Fig.~\ref{param_plane}.\label{determ_trajectory}}
\end{center}
\end{figure}
In the absence of an external driving force there is only one
stationary state given by
\begin{eqnarray}
x_0&=&-I
\label{FitzHugh1_stationary}\\
y_0 &=& -I + I^3/3.
\label{FitzHugh2_stationary}
\end{eqnarray}
The first variable, $x$, represents the membrane potential which is
characterized by a fast dynamics: after $x$ crosses the threshold
value ($x=0$), rapidly it takes on values below the threshold and a
period occurs during which no firing is possible. This condition is
connected with the dynamics of the second variable, $y$,
characterized by a slow dynamics. Depending on the value of $I$ the
stationary point is unstable with stable periodic solution ($|I|<1$)
or stable with all the trajectories converging on it (attractor)
($|I|>1$)~\cite{Pankratova}. Here we set $I=1.1$ and we take
($x_0,y_0$) as initial state for the evolution of the FHN system. In
this condition a time evolution occurs if the system is driven out
of equilibrium. First, we consider the system in deterministic
dynamical regime, that is, when firing events occur in the absence
of noise (deterministic firing region). This regime takes place when
values of amplitude and frequency are chosen inside the gray zone of
the parameter plane (see Fig.~\ref{param_plane}) and corresponds to
the situation shown in Fig.~\ref{determ_trajectory}~(a)). Otherwise,
when the neuron shows no excitability in the absence of noise, the
system is in the deterministic no-firing region (see
Fig.~\ref{determ_trajectory}~(b)). This condition occurs when the
values of $A$ and $\omega$ are chosen out of the gray zone of
Fig.~\ref{param_plane}.

\section{The stochastic FHN model}
\label{sec:stoch_model}

Physical and biological systems are affected by the presence of
continuous random perturbations, due to fluctuations of
environmental parameters such as temperature and natural resources,
which contribute to modify the system dynamics~\cite{Complex}.
%
\begin{figure}
\begin{center}
\resizebox{0.90\columnwidth}{!}{\includegraphics{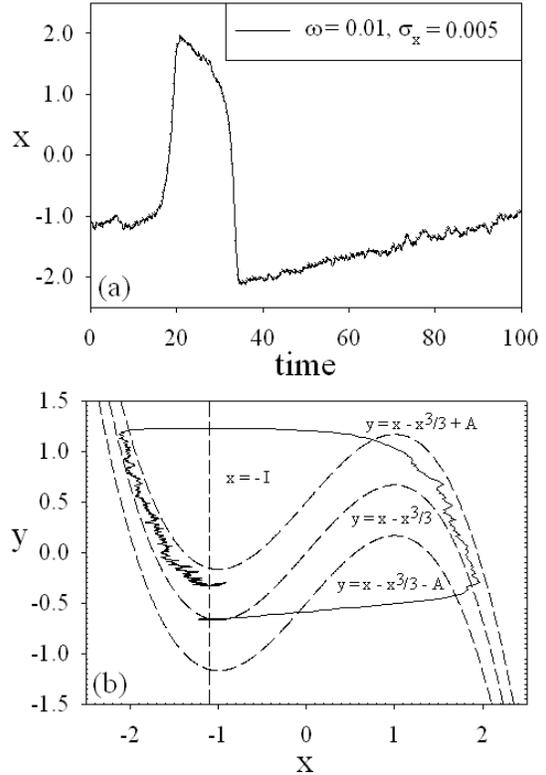}}
\caption{(a) and (b) Stochastic dynamics and corresponding
trajectory in the space ($x,y$) for $A = 0.5$, $\omega = 0.01$:
because of the cooperative action of external signal and
fluctuations a stochastic evolution appears that allows the system
to fire out of the deterministic firing region. Parameter values and
initial conditions are the same of
Fig.~\ref{param_plane}.\label{stoch_trajectory}}
\end{center}
\end{figure}
These fluctuations can be modeled by inserting a noise term in the
deterministic equations. In this work we consider the neuron as an
open system whose dynamics is affected by the presence both of
periodic and random variations of environmental parameters.
Therefore we modify the deterministic FHN model by adding a colored
noise term. In particular we consider two different cases: i) the
membrane potential is subject to a noisy dynamics (Case I); ii) the
refractory variable, which describes the recovery properties of the
neuron, is exposed to fluctuations (Case II). From
Eqs.~(\ref{FitzHugh1}),(\ref{FitzHugh2}) we get
\begin{eqnarray}
Case \enspace I \nonumber\\
\dot{x}&=&x-\frac{x^3}{3}- y + A sin (\omega \thinspace t +
\varphi_0) + \zeta_x \nonumber\\
\dot{y}&=&\epsilon (x+I), \label{FitzHugh_CaseI}\\ \nonumber\\
Case \enspace II \nonumber\\
\dot{x}&=&x-\frac{x^3}{3}- y + A sin (\omega \thinspace t +
\varphi_0)\nonumber\\
\dot{y}&=&\epsilon (x+I) + \zeta_y, \label{FitzHugh_CaseII}
\end{eqnarray}
where $\zeta_i(t)$ ($i = x, y$) are self-correlated noises described
by Ornstein-Uhlenbeck processes~\cite{Gardiner}
\begin{equation}
\frac{d\zeta_i}{dt}=-\frac{1}{\tau}\zeta_i(t)+\frac{1}{\tau}\xi_i(t).
\label{colored_noise}
\end{equation}
Here $\tau$ is the correlation time of the noise, and $\xi_i(t)$ are
statistically independent Gaussian white noises with zero mean and
correlation function
\begin{equation}
<\xi_i(t)\xi_j(t')>=\sqrt{\sigma_i}\thinspace\sqrt{\sigma_j}\thinspace\delta_{ij}
\thinspace \delta(t-t') \quad (i,j =x,y)\label{white_noise}
\end{equation}
The correlation function of the process given by
Eq.~(\ref{colored_noise}) is
\begin{equation}
\langle \zeta_i(t)\zeta_j(t')\rangle =
\frac{\sqrt{\sigma_i}\thinspace\sqrt{\sigma_j}\thinspace}{2 \tau}
e^{-|t-t'|/\tau} \delta_{ij}
 \label{correlation function}
\end{equation}
with
\begin{equation}
lim_{\tau\rightarrow0}\langle \zeta_i(t)\zeta_j(t')\rangle =
\sqrt{\sigma_i}\thinspace\sqrt{\sigma_j}\thinspace\delta(t-t')\delta_{ij}.
 \label{lim_corr_func}
\end{equation}
In a previous work the system response has been studied, under the
influence of a periodic driving signal, as a function of the white
noise intensity~\cite{Pankratova}. The authors showed the occurrence
of resonant activation (RA) and noise enhanced stability (NES). In
particular the presence of a noise source is responsible for
modifications in the resonant activation phenomenon and causes noise
enhanced stability to appear in the minimum. NES effect and
modifications in RA, observed in the deterministic firing region
(grey zone in Fig.2), are connected with the appearance of neuron
excitability even if the parameters $A$ and $\omega$ take values out
of this region. In fact, by setting $A = 0.5$, $\omega = 0.01$ with
$\tau \rightarrow 0$ (white noise) and $\sigma_x = 0.005$ (Case I),
the FHN system can fire also for parameter values chosen in the
deterministic no-firing region (see Fig.~\ref{stoch_trajectory}).
Because of the cooperative action of periodic force and noise, a
stochastic evolution appears that allows the system to fire out of
the deterministic firing region. However, a self-correlated noise
represents a model more suitable to represent the random
fluctuations in real systems, where the noise spectrum is
characterized by the presence of a cut-off. Therefore, in this paper
we study the system response, in the presence of a periodic driving
signal, by using the colored Gaussian noise given by
Eq.~(\ref{colored_noise})~\cite{Valenti}.
%
\begin{figure*}
\begin{center}
\resizebox{1.5\columnwidth}{!}{\includegraphics{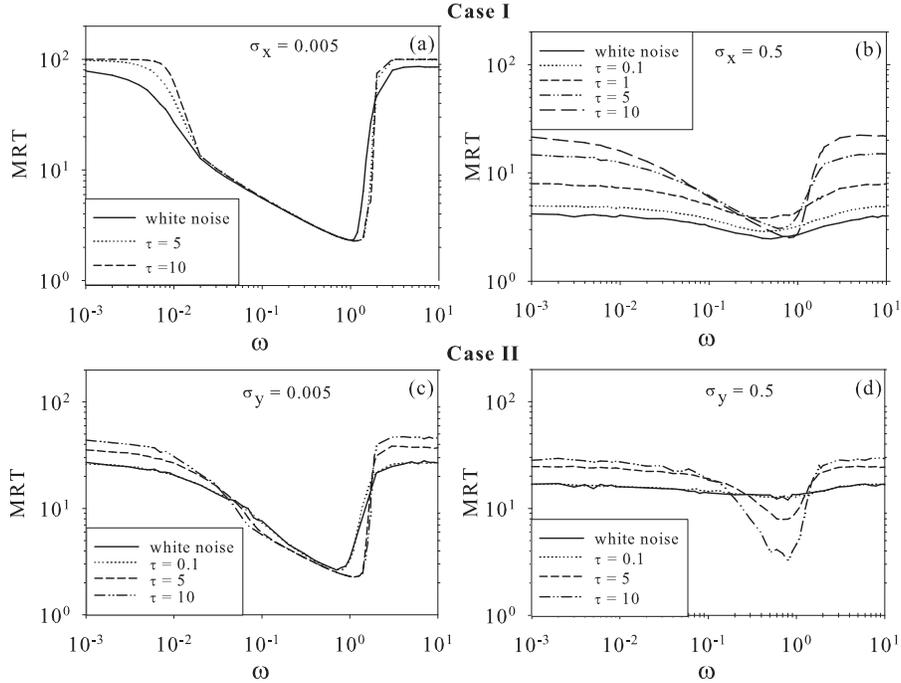}}
\caption{(a) and (c): For low noise intensities a slight
displacement of the RA minimum is observed as the correlation time
$\tau$ increases (weak suppression of the noise effects). (b) and
(d): For higher noise intensities the minimum of RA is strongly
affected by the value of $\tau$. Both in cases I and II the RA
minimum almost disappears for white noise, while it is more
pronounced for high values of $\tau$ (strong suppression of the
noise effects). Parameter values and initial conditions are the same
of Fig.~\ref{param_plane}.\label{RA}}
\end{center}
\end{figure*}

\section{Results}
\label{sec:results}

In order to investigate the dynamics of FHN system we analyze the
behaviour of the mean response time (MRT) of the neuron when
periodical stimulus and colored noise are present. Therefore we
don't consider the periodicity of the signal. MRT is defined as $<T>
\thinspace = \thinspace 1/N \sum_{i=1}^N T_i$. Here $T_i$ is the
\emph{first response time} (that is, the time for which the first
spike occurs) of the $i^{th}$ realization and $N$ is the total
number of realizations. In order to calculate MRT we solve
Eqs.~(\ref{FitzHugh_CaseI},~\ref{FitzHugh_CaseII}) by performing
numerical simulations, with $N=5000$ for case I (the membrane
potential is subject to fluctuations) and $N=15000$ for case II (the
refractory variable is noisy). For all realizations the amplitude
value of the external driving force and the initial conditions are
$A=0.5$ and $(x_0,y_0)$, respectively. We consider a spike occurred
when $x$ gets over the threshold value $x_{thr}=0$. We observe
Resonant Activation (RA) and Noise Enhanced Stability (NES)
phenomena both in cases I and II.

\subsection{Resonant Activation}
\label{sec:RA}

In Fig.~\ref{RA} we report MRT as a function of $\omega$ for
different values both of noise intensities $\sigma_x$, $\sigma_y$
and correlation time $\tau$. The value of the RA minimum is affected
by the noise intensity. For a low level of noise (see
Fig.~\ref{RA}(a),~(c)) this minimum is well pronounced, without any
significative modifications occur as the correlation time $\tau$
increases. We name this \emph{weak suppression of the noise
effects}. However, at higher noise intensities, $\sigma_x = 0.5$ and
$\sigma_y = 0.5$, (see Fig.~\ref{RA}(b),~(d)), the correlation time
becomes more relevant, so that the value of the RA minimum depends
strongly on $\tau$. In particular, as $\tau$ increases, the minimum
is closer to the values of the deterministic regime. In this case we
say that a \emph{strong suppression of the noise effects} occurs. In
particular, in case I we find a nonmonotonic behaviour of MRT as a
function of $\tau$ in the frequency range $0.1<\omega<1.5$.

\subsection{Noise Enhanced Stability}
\label{sec:NES}

By comparing panels (a) and (b) in Fig.~\ref{RA} for $\tau=0$ (white
noise) and $\tau=5,\enspace 10$, we observe, in case I (membrane
potential is subject to fluctuations), an enhancement of the MRT due
to the noise, that is, the depth of RA minimum is reduced as the
noise intensity increases. This enhancement of MRT indicates that
higher levels of noise cause a "response delay" and reduce the
neural efficiency. Extensively we consider this behaviour as NES
effect. A comparison between panels (c) and (d) of Fig.~\ref{RA}
shows that the same phenomenon is present also in case II
(refractory variable is noisy).
%
\begin{figure*}
\begin{center}
\resizebox{1.5\columnwidth}{!}{\includegraphics{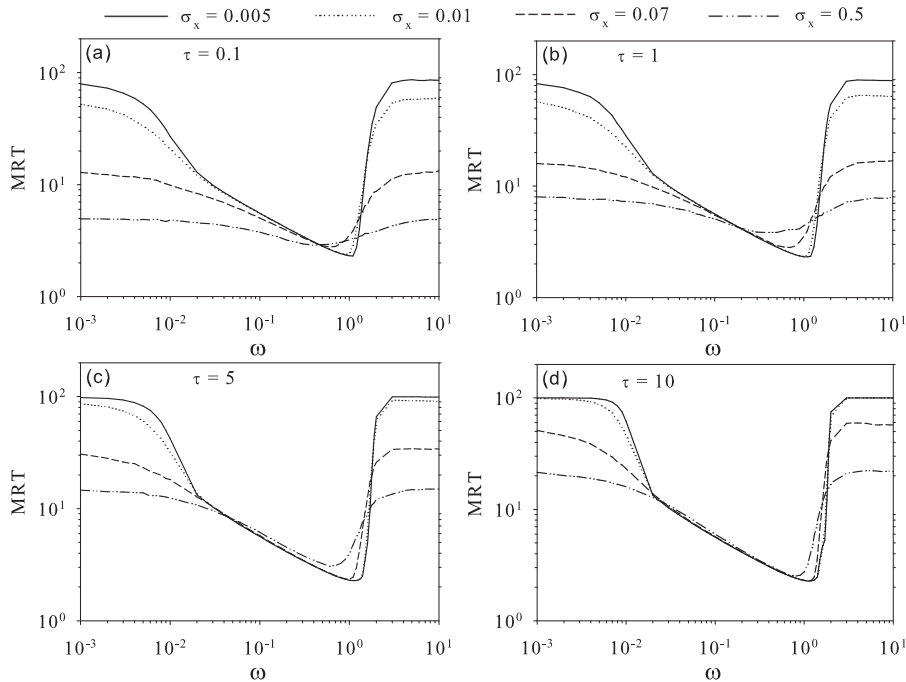}}
\caption{\textbf{Case I}. For weakly correlated noise ($\tau = 0.1$)
MRT shows a nonmonotonic behaviour as a function of the noise
intensity (see panel (a)). For higher values of the correlation time
($\tau = 1$) this effect is reduced (see panel (b)). In the presence
of strongly correlation ($\tau = 5$, $\tau = 10$) this phenomenon
disappears and the depth of RA minimum becomes almost independent on
the noise intensity (suppression of the noise effect)(see panels
(c), (d)). Parameter values and initial conditions are the same of
Fig.~\ref{param_plane}.\label{NES1}}
\end{center}
\end{figure*}
%
\begin{figure*}
\begin{center}
\resizebox{1.5\columnwidth}{!}{\includegraphics{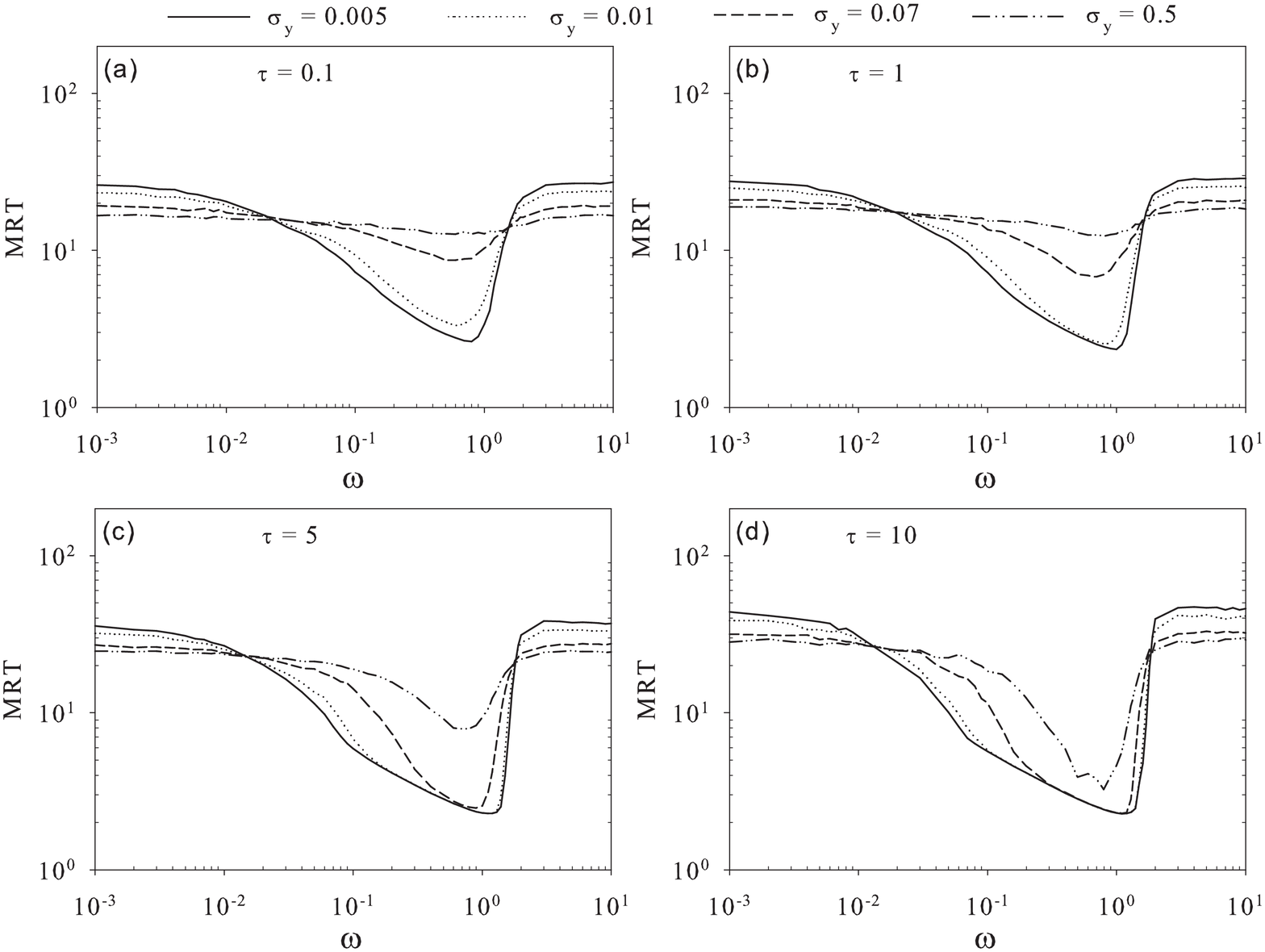}}
\caption{\textbf{Case II}. For weakly correlated noise ($\tau =
0.1$) a monotonic behaviour of MRT as a function of the noise
intensity appears (see panel (a)). As the correlation time increases
($\tau = 1$) this phenomenon is attenuated (see panel (b)). Finally,
for strongly correlation ($\tau = 5$, $\tau = 10$) the depth of RA
minimum is almost independent on the noise intensity (suppression of
the noise effect). Parameter values and initial conditions are the
same of Fig.~\ref{param_plane}.\label{NES2}}
\end{center}
\end{figure*}
In view of a better understanding of this effect, we analyze further
the behavior of the mean response time. In particular, we calculate
the MRT as a function of the noise intensity for different values of
the correlation time. The results are shown in
Figs.~\ref{NES1},~\ref{NES2}. We note that a weakly correlated noise
source affects significatively the efficiency of the neuronal
response. In particular, for $\tau=0.1, 1$ we observe that, both in
cases I and II, the depth of RA minimum decreases for high levels of
noise, which is an enhancement of MRT due to the noise (NES effect)
(see Figs.~\ref{NES1}a,b and \ref{NES2}a,b). Conversely, in the
presence of strongly correlated noise, no lack of efficiency is
observed around the RA minimum, which maintains its depth almost
unchanged as noise intensity increases (see Figs.~\ref{NES1}c,d and
\ref{NES2}d). This behavior can be evidenced by taking into account
the values of MRT in the RA minimum. From Figs.~\ref{NES1} and
\ref{NES2} we note that the position of the minimum, for the lowest
values of the correlation time and the noise intensity ($\tau = 0.1$
and $\sigma=0.005$), is given by $\omega \simeq 1$ for case I and
$\omega \simeq 0.7$ for case II. Therefore we calculate MRT as a
function of the noise intensity for different values of $\tau$,
setting $\omega=1$ in case I and $\omega=0.7$ in case II. The
results are reported in Fig.~\ref{NES3}.
%
\begin{figure}
\begin{center}
\resizebox{0.85\columnwidth}{!}{\includegraphics{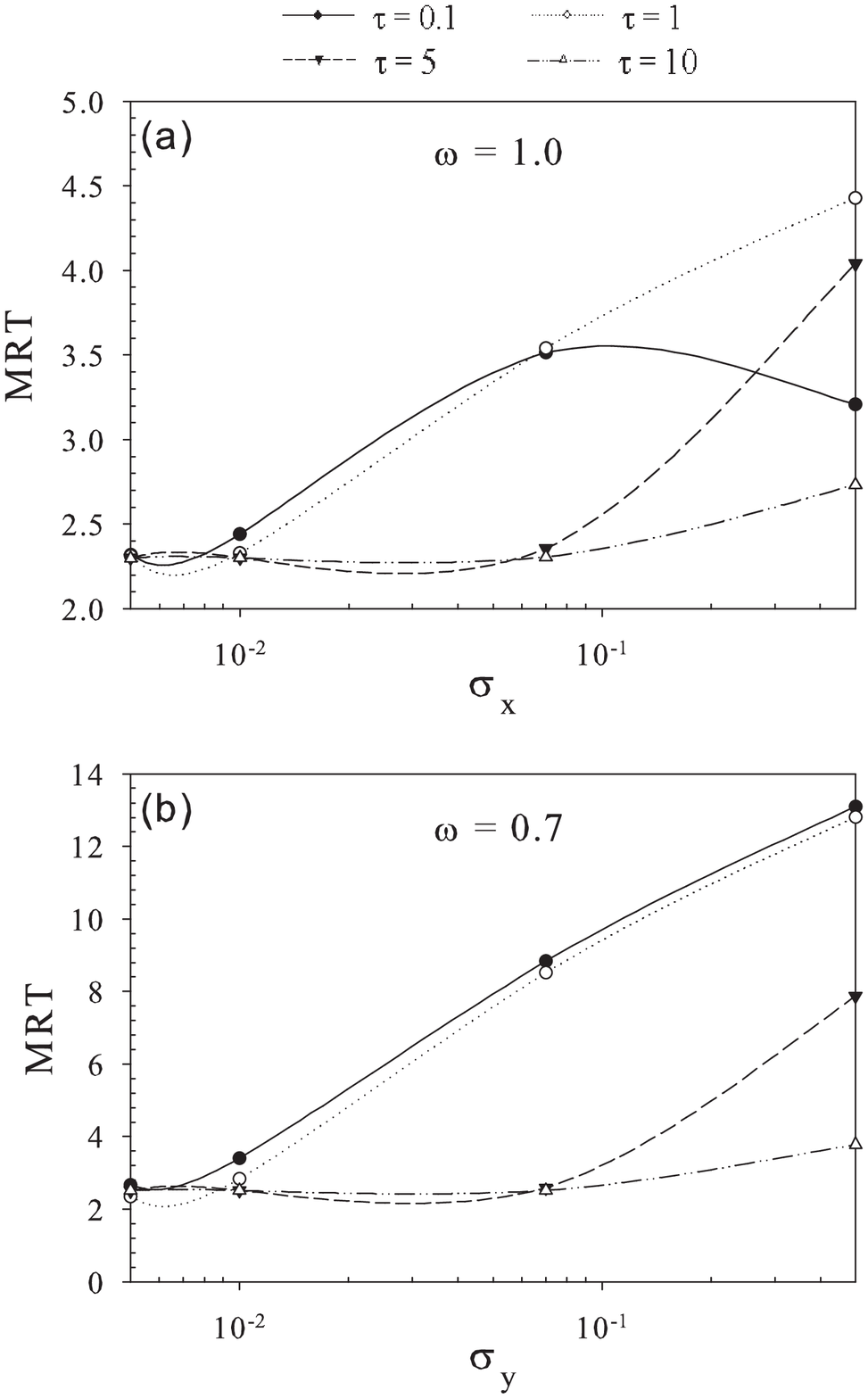}}
\caption{(a) Case I. Nonmonotonic behaviour (with a maximum) of MRT
as a function of the noise intensity $\sigma_x$, for $\omega=1.0$:
in the presence of colored noise this effect disappears as
correlation time increases. (b) Case II. Monotonic increase of MRT
as a function of the noise intensity $\sigma_y$, for $\omega=0.7$:
this behaviour is suppressed for strongly correlated noise.
Parameter values and initial conditions are the same of
Fig.~\ref{param_plane}.\label{NES3}}
\end{center}
\end{figure}
\begin{figure}
\begin{center}
\resizebox{0.85\columnwidth}{!}{\includegraphics{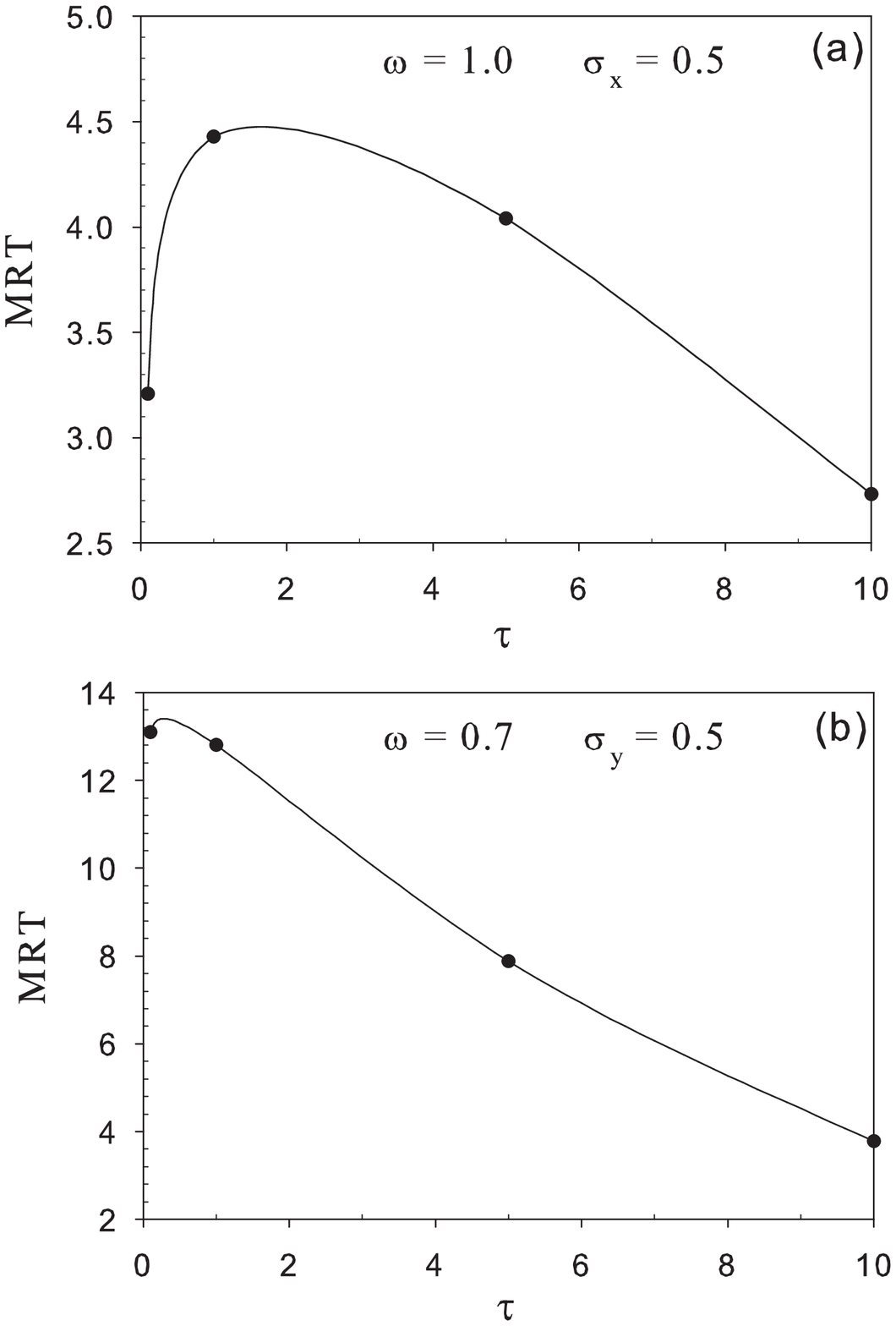}}
\caption{(a) Case I. Nonmonotonic behaviour of the MRT as a function
of the correlation time $\tau$, for $\omega=1.0$ and $\sigma_x=0.5$.
(b) Case II. Decrease of the MRT as a function of the correlation
time $\tau$, for $\omega=0.7$ and $\sigma_y=0.5$. Parameter values
and initial conditions are the same of
Fig.~\ref{param_plane}.\label{MRT_vs_tau}}
\end{center}
\end{figure}
From the inspection of the figure we note that, in both cases, the
strong dependence of the MRT on the noise intensity is suppressed as
correlation time increases: the effect of the noise in the RA
minimum (region of highest efficiency) appears to be smaller as
$\tau$ increases. More precisely we note that, in case I, MRT shows
in the minimum a nonmonotonic behavior as a function of $\sigma_x$:
starting from $\sigma_x=0.005$ for intermediate levels of noise
intensity an increase of MRT appears. For higher noise intensities a
decrease of MRT is observed (see Fig.~\ref{NES3}a). On the other
side, for case II, a monotonic increase of MRT is observed as a
function of the noise intensity $\sigma_y$. However, in both cases,
for high values of the correlation time, MRT tends to be constant as
the noise intensity increases, in the range of the noise intensity
values investigated. The modifications induced by the noise in the
mean response time are strongly reduced and they almost disappears
for $\tau=10$ (suppression of the noise effects). Finally we note,
in case I, a nonmonotonic behaviour of MRT as a function of the
correlation time $\tau$. This behaviour appears for noise intensity
values greater than $\sigma_x=0.07$, corresponding to the maximum of
the curve for $\tau = 0.1$. In Fig.~\ref{MRT_vs_tau} we report MRT
vs $\tau$ both for case I and II with $\sigma = 0.5$.

\subsection{Colored noise: rescaling effect}
\label{subsec:rescaling}

Let's consider two colored Gaussian noise sources, $S1$ and $S2$,
with the same intensity $\sigma$, characterized by the correlation
times $\tau_1$ and $\tau_2$ respectively, with $\tau_1<\tau_2$. The
behaviour shown in Fig.~\ref{NES3} indicates that the presence of a
self-correlation causes a reduction of the noise "efficacy". This
suggests that, for frequency values around the RA minimum, the noise
effect on a FHN system is more intensive when one uses $S1$, that is
the noise source characterized by the smaller correlation time.
Remembering that the white noise is obtained by the colored noise in
the limit $\tau \rightarrow0$ (see Eq.~(\ref{lim_corr_func})), one
expects that, for frequency values around the RA minimum, a white
noise source with intensity $\sigma$ influences the neural dynamics
more than any correlated noise source with the same intensity
$\sigma$.

\begin{figure}
\begin{center}
\resizebox{0.85\columnwidth}{!}{\includegraphics{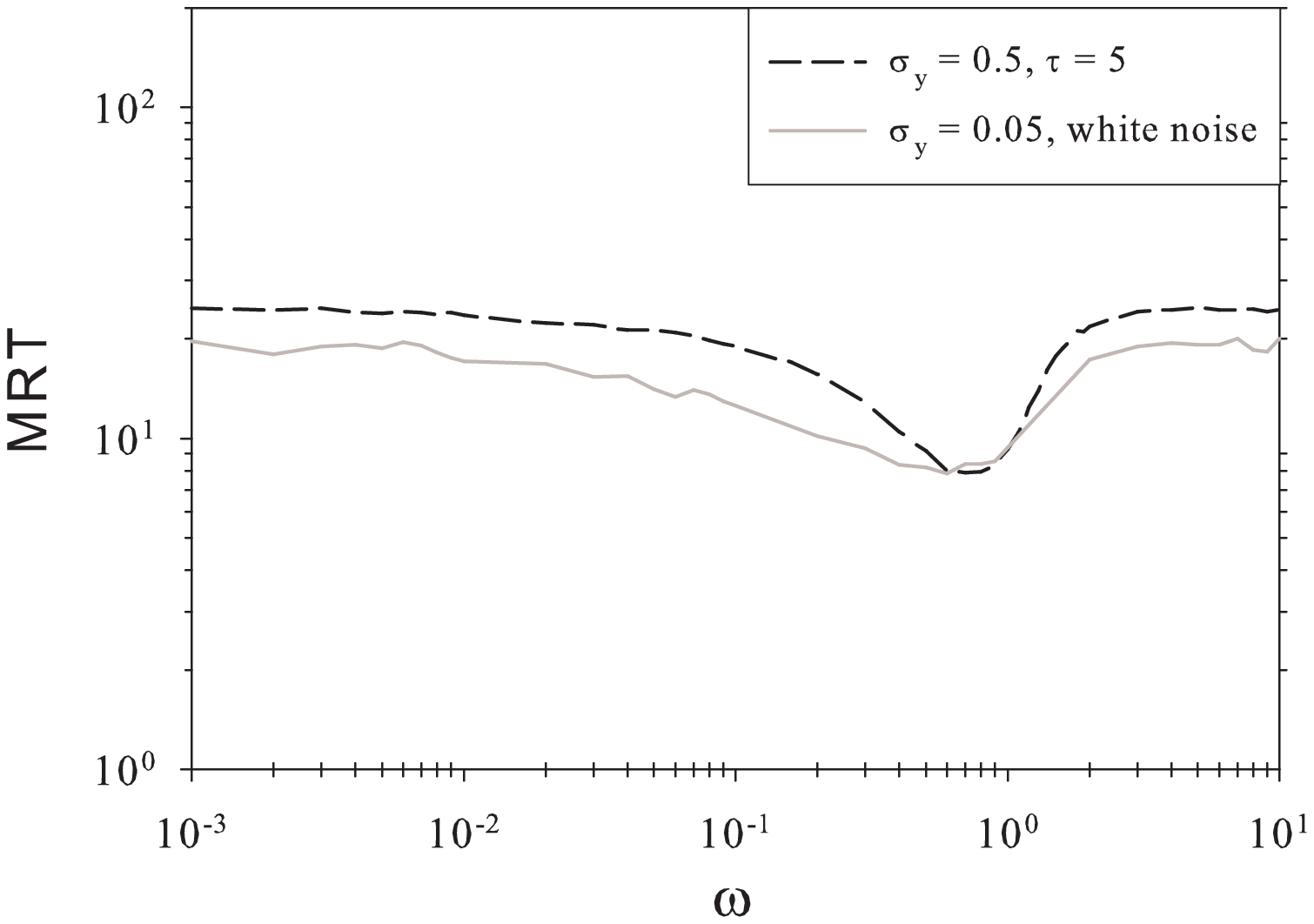}}
\caption{Case II. The correlation time causes a rescaling effect
according to $\sigma_{white} = \sigma_{colored}/2\tau$. Around the
minimum the curve obtained for $\sigma_{colored} = 0.5$ and
correlation time $\tau = 5$ (black dash line) overlaps that obtained
by using a white noise source with intensity
$\sigma_{white}=\sigma_{colored}/2\tau=0.05$ (gray solid line).
Parameter values and initial conditions are the same of
Fig.~\ref{param_plane}.\vspace{1mm}\label{rescaling}}
\end{center}
\end{figure}

From Eq.~(\ref{lim_corr_func}) we note that for $\vert t-t'\vert <
\tau$ we can approximate $e^{-|t-t'|/\tau}\sim 1$. So that from
Eq.~(\ref{correlation function}) we get
\begin{equation}
\langle \zeta_i(t)\zeta_j(t')\rangle \approx
\frac{\sqrt{\sigma_i}\thinspace\sqrt{\sigma_j}\thinspace}{2 \tau}
\delta_{ij}.
 \label{new_correlation function}
\end{equation}
This result could be interpreted as a rescaling effect: the
correlation time seems to reduce the noise intensity of the colored
noise by a factor equal to $2\tau$. This suggests that, in the RA
minimum, the effect of a colored noise source with intensity
$\sigma_{colored}$ and correlation time $\tau$ should be, for $\vert
t-t'\vert/\tau \sim 0$, the same obtained by using a white noise
source with intensity given by
\begin{equation}
\sigma_{white} = \frac{\sigma_{colored}}{2\tau}.
 \label{new_correlation function}
\end{equation}
Therefore we define the "effective" intensity of a colored noise
according to Eq.~(\ref{new_correlation function}), when MRT is of
the same order of magnitude of $\tau$ or less. In order to verify
this rescaling effect we consider again the curve of Fig.~\ref{RA}d
obtained for $\sigma_y=0.5$ and $\tau = 5$ and we obtain the
corresponding "effective" noise intensity $\sigma_{white} =
(\sigma_{colored})/2\tau = 0.05$. Afterwards, by using this value we
calculate the MRT as a function of $\omega$. The results have been
reported in Fig.~\ref{rescaling}.
%
In the figure we note that around the RA minimum, that is for $MRT
\sim \tau$ and $e^{-|t-t'|/\tau}\sim 1$, the influence of the
colored noise in the FHN dynamics depends on the "effective" noise
intensity $\sigma_{white}=0.05$. This means that, in the presence of
colored noise and for suitable values of the correlation time
$\tau$, the actual noise intensity perceived by the FHN system is
reduced by the rescaling factor $1/(2\tau)$. We recall that the
white noise is a theoretical assumption to describe noise whose band
width is very large. In fact, in real systems the fluctuations are
connected with colored noise sources. Therefore, in order to
evaluate the response of a real neuron, one has to consider
self-correlated noise sources: the rescaling phenomenon, as we show
in this paper, reduces the effect of the noise on the FHN dynamics,
modifying in a significative way the neuronal dynamics.

\section{Conclusions}
\label{sec:conclusions}

After a brief discussion on the ~Hodgkin-Huxley (HH) model, we
introduce the FitzHugh-Nagumo (FHN) model. Although it is a
simplified version of the former, the latter permits to separate the
mathematical characteristics of excitation and propagation from the
electrochemical peculiarities of sodium and potassium ion flows.
Even if the HH model better describes the real dynamics of the
neuronal response, it allows only to observe two-dimensional
projections of its four-dimensional phase trajectories. Because of
this, the FHN model is often preferred, providing the whole solution
directly in a two-dimensional phase space. This characteristic
permits to envisage a geometrical interpretation of important
biological phenomena that depend on the spike-generating mechanism
which causes the neuronal response to external stimuli. Therefore,
in this paper we analyzed the response of a neuron in the presence
both of a driving periodical force with frequency $\omega$ and an
additive Gaussian colored noise, by the FHN neuronal model. In our
analysis we considered two cases: I) the noise source affects the
membrane potential, II) the noise source influences the recovery
variable. In these conditions we analyzed the mean response time
(MRT) as a function both of the noise intensity and the correlation
time. We found that RA and NES phenomena undergo meaningful
modifications due to the presence of different values of the
correlation time $\tau$. For strongly correlated noise we observed
suppression of NES and persistence of RA (efficiency enhancement of
neuronal response) as correlation time $\tau$ increases. This
indicates that the RA minimum corresponding to a certain value of
$\omega$ is preserved when the noise source is characterized by
large values of the correlation time. Conversely the enhancement of
the MRT, which indicates a significant role of the noise, tends to
vanish. The reduction of the noise effects in the presence of
strongly correlated noise indicates a rescaling effect of the noise
self-correlation. We investigated this aspect for a colored noise
source whose intensity and correlation time are given by $\sigma$
and $\tau$ respectively. We found that, for $MRT\sim \tau$, an
"effective" noise intensity exists. This implies that the effect of
the self-correlated noise can be reproduced, through a rescaling
procedure, by using a white noise source whose amplitude is given by
$\sigma_{white} = (\sigma_{colored})/2\tau$.

\section{Acknowledgments}
\label{sec:acknowledgments}

Authors acknowledge the support by MUR. This work makes use of
results produced by the PI2S2 Project managed by the Consorzio
COMETA, a project co-funded by the Italian Ministry of University
and Research (MIUR) within the Piano Operativo Nazionale "Ricerca
Scientifica, Sviluppo Tecnologico, Alta Formazione" (PON 2000-2006).
More information is available at http://www.pi2s2.it and
http://www.consorzio-cometa.it.

\end{document}